\documentclass[journal=jacsat,manuscript=article]{achemso}

\usepackage{chemformula} % Formula subscripts using \ch{}
\usepackage[T1]{fontenc} % Use modern font encodings
\usepackage{graphicx} % To import figures
\usepackage{xcolor}
\author{Mahmoud R. M. Atalla, C\'edric Lemieux‐Leduc, Simone Assali, Sebastian Koelling, Patrick Daoust, and Oussama Moutanabbir}
\email{ oussama.moutanabbir@polymtl.ca} %% email address is required; see note below about the corresponding author designation
\affiliation{Department of Engineering Physics, \'Ecole Polytechnique de Montr\'eal, C.P. 6079, Succ. Centre-Ville, Montr\'eal, Qu\'ebec, Canada H3C 3A7}
\title {Extended-SWIR High-Speed All-GeSn PIN Photodetectors on Silicon   }

\begin{document}

\begin{abstract}
\noindent There is an increasing need for silicon-compatible high bandwidth extended-short wave infrared (e-SWIR) photodetectors (PDs) to implement cost-effective and scalable optoelectronic devices. These systems are quintessential to address several technological bottlenecks in detection and ranging, surveillance, ultrafast spectroscopy,  and imaging. In fact, current e-SWIR high bandwidth PDs are predominantly made of III-V compound semiconductors and thus are costly and suffer a limited integration on silicon besides a low responsivity at wavelengths exceeding $2.3 \,\mu$m. To circumvent these challenges, Ge$_{1-x}$Sn$_{x}$ semiconductors have been proposed as building blocks for silicon-integrated high-speed e-SWIR devices.  Herein, this study demonstrates a vertical all-GeSn PIN PDs consisting of p-Ge$_{0.92}$Sn$_{0.08}$/i-Ge$_{0.91}$Sn$_{0.09}$/n-Ge$_{0.89}$Sn$_{0.11}$ and p-Ge$_{0.91}$Sn$_{0.09}$/i-Ge$_{0.88}$Sn$_{0.12}$/n-Ge$_{0.87}$Sn$_{0.13}$ heterostructures grown on silicon following a step-graded temperature-controlled epitaxy protocol. The performance of these PDs was investigated as a function of the device diameter in the $10-30 \,\mu$m range. The developed PD devices yield a high bandwidth of 12.4 GHz at a bias of 5V for a device diameter of $10 \,\mu$m. Moreover, these devices  show a high responsivity of 0.24 A/W, a low noise, and a $2.8 \,\mu$m cutoff wavelength thus covering the whole e-SWIR range.

\end{abstract}

\section{Introduction}

% Paper structure 52 rf:

Extended-short wave infrared (e-SWIR) photodetectors (PDs) featuring high bandwidth and wide spectrum range are critical to a plethora of applications spanning free-space and fiber-coupled communications, high temporal resolution light detection and ranging (LIDAR), environmental gas sensing, and time resolved spectroscopy.\cite{wun2016,Williams2017,Kim2021, wang2021high2,li202130,razeghi2014advances, sieger2016toward, lin2018mid, popa2019towards, chen2021recent} This strategic range of the electromagnetic spectrum is currently predominately served by compound semiconductors. For instance, the III-V e-SWIR PDs, typically grown on InP or GaSb substrates, can operate at wavelengths exceeding $2.3 \,\mu$m. \cite{Ye2015,joshi2008high,yang2013,chen2018,tossoun2018,chen2019hi,andreev2013high,wun2016} These devices are, however, prone to a limited bandwidth which remains below 6 GHz. Although the development of GeSn e-SWIR photodetectors is still in its infancy, these semiconductors offer a viable alternative to circumvent the current limitations in both wavelength range and operation speed in addition to their compatibility with silicon-based complementary metal-oxide-semiconductor (CMOS) technology. Indeed, GeSn is an all-group IV alloy with a demonstrated content-dependent tunable bandgap energy covering the entire IR spectrum.\cite{moutanabbir2021mono,buca2022room,atalla2023extended,chang2022mid,chretien2022room,elbaz2020ultra,jung2022optically,li202130,liu2022sn,luo2022extended,marzban2022strain,talamas2021cmos,daligou2023group,mircovich2021extended,xu2019synthesis,soref2015enabling,elbaz2020ultra,zhou2020elec,chtien2019} 

GeSn PIN PDs can be either free-space or waveguide devices that are directly integrated on Si substrates.\cite{xu2019high,tran2019study,talamas2021,tsai2021gesn,wang2021high2,li202130} Free-space  GeSn/Ge PDs with a Sn composition gradient in the active layer have recently shown a bandwidth of 50 GHz and a 2.8 $\,\mu$m cut-off, which highlights the potential of this material system.\cite{cui2023sn} Notwithstanding this progress, the responsivity of these devices drops significantly at wavelengths above 2.3 $\,\mu$m owing to the remarkable Sn composition gradient and high residual compressive strain that is typical to the GeSn/Ge heterostructures used to implement these PDs. In contrast, the growth of all-GeSn PIN heterostructures can yield thicker yet more relaxed GeSn active layers with uniform Sn composition \cite{atalla2023extended} . These characteristics are needed to achieve a high and uniform responsivity across a larger wavelength range\cite{tran2019si}. Up to date, the bandwidth of the demonstrated all-GeSn e-SWIR PDs operating at wavelength exceeding $2.3 \,\mu$m are limited to 7.5 GHz at $2.6 \,\mu$m cutoff wavelength.\cite{atalla2022high} Herein, this work demonstrates PIN GeSn photodetectors of small active area diameters below $30 \,\mu$m. These devices are made of  vertical all-GeSn PIN PDs consisting of p-Ge$_{0.92}$Sn$_{0.08}$/i-Ge$_{0.91}$Sn$_{0.09}$/n-Ge$_{0.89}$Sn$_{0.11}$ and p-Ge$_{0.91}$Sn$_{0.09}$/i-Ge$_{0.88}$Sn$_{0.12}$/n-Ge$_{0.87}$Sn$_{0.13}$ heterostructures grown on silicon following a step-graded process. It is shown that 12.4 GHz operation at room temperature for a bias of 5V is achievable by reducing the device diameter down to $10 \,\mu$m to minimize the total capacitance. The normalized response curves of the PIN devices indicate that the bandwidth is still RC limited. It is also shown that these devices exhibit a responsivity of 0.24 A/W at $1.55 \,\mu$m and cutoff wavelength up to $2.8 \,\mu$m. The responsivity of these devices does not degrade significantly at wavelengths above 2.3 $\,\mu$m owing to the relaxed all-GeSn heterostructures.

\section{Results and discussion}

\noindent {\bf Growth and characterization of GeSn epilayers.} 

\noindent The all-GeSn PIN heterostructures were grown in a low-pressure chemical vapor deposition (CVD) reactor on a 4-inch Si ($100$) wafer using Ge virtual substrate (Ge-VS). The growth protocol\cite{atalla2022high} involves the growth of multiple GeSn buffer layers with increasing Sn content to accommodate the lattice-mismatch between the device layer and the Ge-VS, followed by PIN growth on top.  The two sets of samples were grown at 305 $^{\circ}$C with the i-GeSn layer composition of 9 at.\% (sample A) and 12 at.\% (sample B) . The higher Sn content in sample B was obtained by increasing the SnCl$_4$ flow by 40\% as compared to sample A. Fig. 1 displays a comparison of the structural properties of the as-grown PIN GeSn stacks. The cross-sectional transmission electron micrographs (TEM) in Fig. 1(a) and 1(b) show that dislocations extend to p-GeSn layer, while leaving the i-GeSn and n-GeSn regions with higher crystalline quality. Diborane (B$_{2}$H$_{6}$) and arsine (AsH$_{3}$) precursors were used to achieve the p- and n-type doping, respectively. The p- and n-layers had carrier concentrations exceeding $1 \times 10^{19}$ cm$^{-3}$ as determined by capacitance-voltage (C-V) measurements. The composition and strain of the as-grown heterostructures were determined using X-ray diffraction (XRD) Reciprocal Space Mapping (RSM) measurements around the asymmetrical (-2-24) reflection peak. The obtained maps for the 9 at.\% and 12 at.\% samples are shown in Figs. 2(c) and 2(d), respectively. Both samples exhibit a low residual compressive strain of $-\,0.2\,\%$ in the buffer layers \#1-3, while this value slightly increases to $-\,0.3\,\%$ in the GeSn  \#4 layer. For sample A, the measured residual strain for the PIN layers is $-\,0.11\,\%$ (p-Ge$_{0.92}$Sn$_{0.08}$), $-\,0.28\,\%$ (i-Ge$_{0.91}$Sn$_{0.09}$), and $-\,0.48\,\%$ (n-Ge$_{0.89}$Sn$_{0.11}$), while for sample B the strain values are $-\,0.19\,\%$ (p-Ge$_{0.91}$Sn$_{0.09}$), $-\,0.36\,\%$ (i-Ge$_{0.88}$Sn$_{0.12}$), and $-\,0.51\,\%$ (n-Ge$_{0.87}$Sn$_{0.13}$).

\begin{figure*}[t]
    \centering
    \includegraphics[scale=0.77]{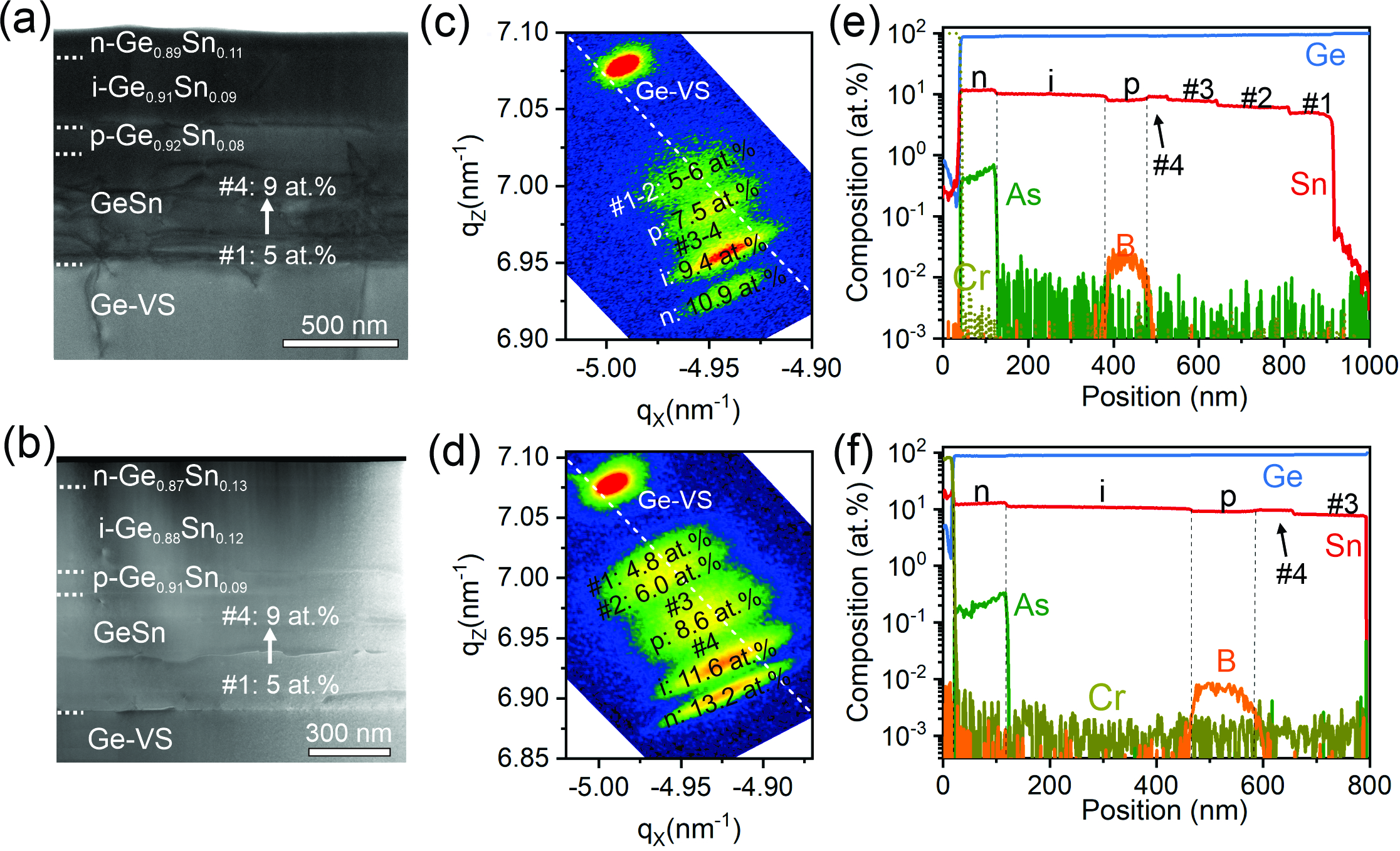}
    \caption{{\bf Structural properties of as-grown GeSn PIN heterostructures.} (a) and (b) Cross-sectional TEM images of sample A (p-Ge$_{0.92}$Sn$_{0.08}$/i-Ge$_{0.91}$Sn$_{0.09}$/n-Ge$_{0.89}$Sn$_{0.11}$) and sample B (p-Ge$_{0.91}$Sn$_{0.09}$/i-Ge$_{0.88}$Sn$_{0.12}$/n-Ge$_{0.87}$Sn$_{0.13}$), respectively. (c) and (d) RSM (-2-24) maps for both samples, respectively. (e) and (f) 3D atom-by-atom reconstructions of the atom probe tips showing the elemental distribution of the As, B, Ge, Sn and Cr atoms for both samples, respectively. Panel (e) was adapted from Ref. [\cite{atalla2022high}].}
\end{figure*}

\medskip

Furthermore, atom probe tomography (APT) was utilized to map the composition and doping profiles in the as-grown heterostructures down to the atomic level. To facilitate the preparation of the APT tips, chromium (Cr) was deposited on top of the as-grown samples using e-beam evaporation before the focused ion beam processing of APT specimens. The obtained APT elemental profiles are shown in Figs. 1(e) and 1(f). Starting by the Sn profile, the uniformity of the Sn composition in the GeSn buffer layers and the PIN stack is evident in both samples. The step-wise increase in the Sn content in the four GeSn buffer layers was achieved by reducing the growth temperature in steps of $10\,^\circ\text{C}$ during the epitaxial growth. This reduction in temperature allows more Sn to be incorporated in the lattice and produces sharp interfaces without any composition gradient. However, the p-GeSn layer exhibits a wider interface owing to the slow incorporation of B dopants in the GeSn lattice as evidenced by the gradually increasing B concentration in the p-GeSn layer. The n-GeSn is doped by As, however, a careful analysis had to be carried out to decouple the signal of As in APT mass spectra from that of Ge due to the small difference between their atomic masses. Additionally, it is inferred from Figs. 1(e) and 1(f) that the interface between n-/i-GeSn layers is abrupt and that the Sn content increases as the As dopants are easily incorporated into the lattice of the growing GeSn layer in sharp contrast to the case of B dopants. It is worthy to mention that sample A has a higher As and B concentrations as compared to sample B. For instance, the As atomic composition is $\sim \, 0.6 $ at.\% in the n-layer of sample A compared to just $\sim \, 0.3 $ at.\% in sample B.  Additionally, the B atomic composition is $\sim \, 0.03 $ at.\% in the p-layer of sample A compered to just $\sim \, 0.01 $ at.\% in sample B. This reduction in the As and B dopant concentrations in sample B is mostly likely related to the increased supply of Sn during the growth of this sample.

\noindent {\bf GeSn PIN photodetectors}

\noindent Samples A and B were subsequently used to fabrication the PIN PDs following a top-down microfabrication process. The device processing started by an ICP Cl$_2$ etch down to Ge-VS to a depth of $\sim 950$ nm followed by a second ICP Cl$_2$ etch of circular bumps of various diameters down to the p-GeSn layer, which forms a double mesa structure to help isolate every device from its neighbouring ones. Then, wet chemical passivation was made followed by PECVD deposition of $1.5\,\mu$m-thick SiO$_2$ as an insulation layer. Next, openings in the SiO$_2$ layer atop the p- and n-GeSn regions were created in a double-step etch consisting of ICP dry etch of 1100 nm followed by wet BOE etch of 400 nm. Finally, a 700 nm-thick Al metal was sputtered followed by contacts patterning and Al wet etch. A scanning electron micrograph (SEM) of a representative $10\,\mu$m-diameter device is shown in Fig. 2(a).

The processed PIN devices were then subject to various sets of electrical and optoelectronic analyses. The recorded I-V curves without illumination are displayed in Fig. 2(b) for both samples A and B for devices with diameters $10$, $20$, and $30 \,\mu$m. It is noticeable that sample A has a relatively high rectification ratio around an order of magnitude at 1V, whereas sample B exhibits very low rectification ratio approaching unity at the same voltage. This can be attributed to the high defect density in the PIN stack of sample B as compared to sample A, which significantly increases the leakage current as the reverse bias increases. In addition, the relatively lower As composition (Figs. 1(e) and 1(f)) in the n-GeSn of sample B compared to sample A explains the reduced forward bias current of sample B at 1V. Note that the dark current, which is the PD current under no illumination, increases monotonically as the device diameter increases for both samples in reverse bias. This can be attributed to bulk and/or surface leakage current. In Fig. 2(c), the dark current density is plotted as a function of applied bias, and it shows that it is almost independent of device diameter. This indicates a low surface leakage current and suggests that the bulk leakage in sample B is most likely the reason underlying its high reverse leakage current.

\medskip

\begin{figure*}[t]
    \centering
    \includegraphics[scale=0.77]{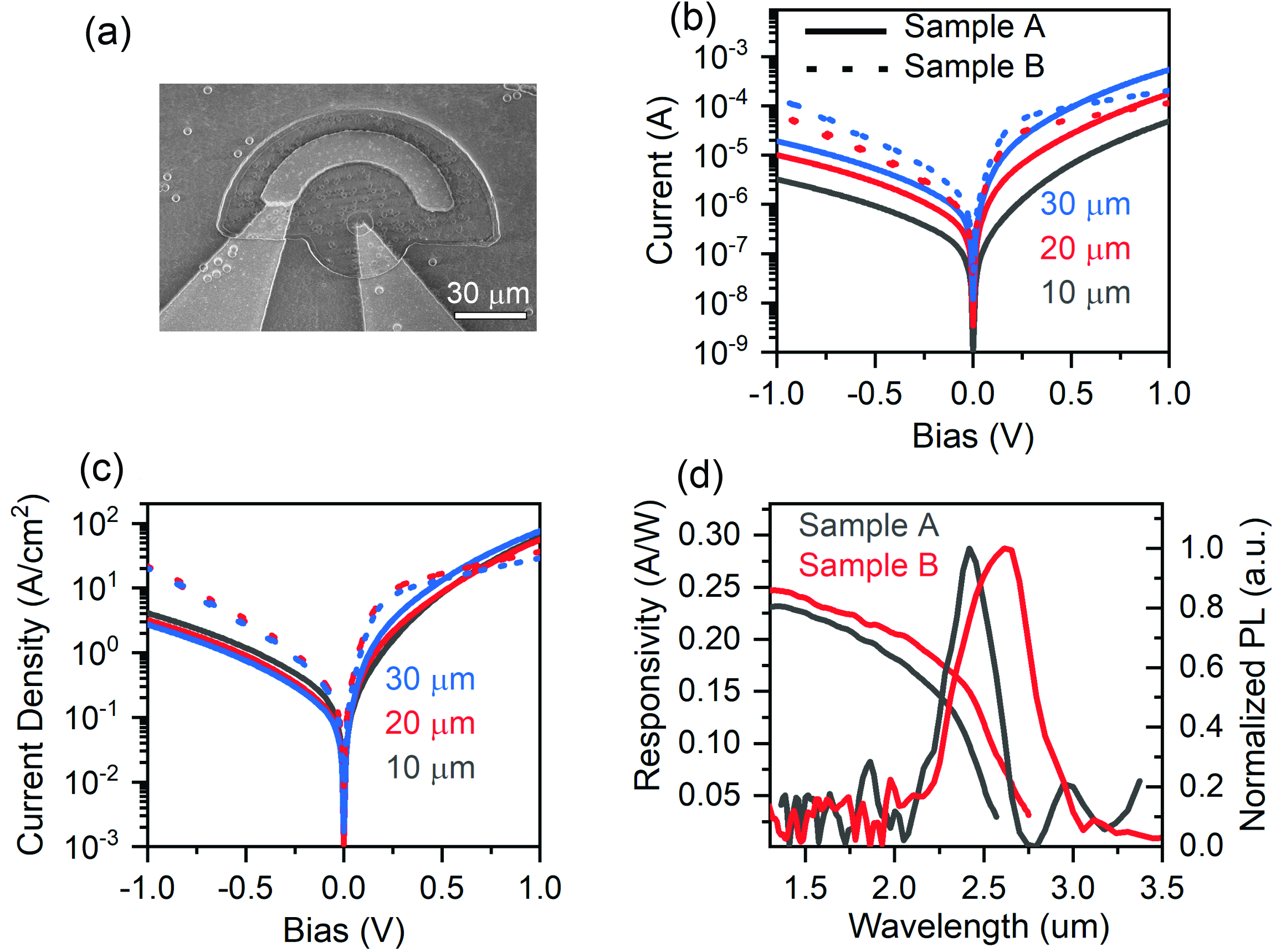}
    \caption{{\bf GeSn PIN photodetectors fabrication and characterization.} (a) SEM micrograph of a representative PIN PD with $10 \,\mu$m in diameter. (b) Room-temperature I-V curves of dark current at various device diameters of sample A compared to sample B. (c) same as (b) but for the dark current density. (d) Spectral responsivity of sample A compared to that of sample B along with the measured room-temperature PL for both samples.}
\end{figure*}

To investigate the device performance under illumination, Fourier transform infrared (FTIR) spectrometer was used to measure the relative spectral responsivity of the PIN GeSn PDs for both samples. The absolute spectral responsivity can be estimated by using a calibrated e-SWIR InGaAs PD . The obtained spectral responsivities of the PIN PDs are plotted in Fig. 2(d) as a function of the incident illumination wavelength. Responsivities of $0.221$ A/W and $0.235$ A/W were measured at $1.55 \,\mu$m for both samples A and B, respectively. In both cases, a monotonic decrease in the responsivity is visible until their spectral responsivity reaches cut-off at $2.6 \,\mu$m (sample A) and $2.8 \,\mu$m (sample B). It is important to note that the all-GeSn heterostructures are characterized by a uniform high Sn content in the i-layers thus yielding a high responsivity that reduces only close to the cutoff wavelength. This is a clear advantage of the all-GeSn PIN PDs as compared to GeSn/Ge heterostructures that can suffer from high residual strain and Sn composition gradient, which limits the responsivity as the wavelength increases. The FTIR spectrometer is also used to measure the photoluminescence (PL) spectra of the as-grown samples A and B as displayed in Fig. 2(d). The PL spectra of samples A and B show peaks at $2.4 \,\mu$m and $2.65 \,\mu$m and their decay extend beyond $2.6 \,\mu$m and $2.8 \,\mu$m, respectively. This is consistent with the estimated cut-off wavelength values mentioned previously. It is important to note that the high responsivity and low dark current increase both the PD sensitivity and signal to noise ratio, which in turn voids the need for a lock-in technique to extract the photocurrent\cite{atalla2021all,tran2019si,atalla2022high}.

\medskip

\noindent {\bf Photoresponse bandwidth of GeSn PDs}

\noindent There are several factors that effect the photoresponse bandwidth of a PD device. In PIN devices, photocarriers generated outside the depletion region in the n- and p-layers transport to the contacts by diffusion. This diffusion current can slow down the device photoresponse. Owing to the remarkable short lifetime of photocarriers in GeSn, the diffusion current is most likely too small and shall be neglected in the calculation of the photoresponse bandwidth. Additionally, photocarrier trapping and release at the heterojunction interfaces produces a slow photoreponse component, however, this component also reduces as the applied reverse bias increases and shall be neglected as well. Finally, the photocarrier transit lifetime through the depletion region and the resistance capacitance (RC) delay are considered the main components limiting the photoresponse bandwidth. Consequently, the $3$\,dB bandwidth, $f_{-3\,\text{dB}}$, can be written as:\cite{dong2017two,lin2017high}  $f_{total}={(f_{RC}^{-2}+f_{trans}^{-2})}^{-0.5}$, where $f_{RC}=1/(2\pi\text{RC})$ is the RC-limited bandwidth and R and C are the total resistance and the total capacitance, respectively. The total resistance includes the contacts resistance, semiconductor resistance, series resistance and load resistance, while the total capacitance includes the contacts parasitic capacitance and depletion region junction capacitance. $f_{trans}=0.45\, v_s/d$ is the transit lifetime-limited bandwidth, where $v_s$ is the saturation velocity and $d$ is the i-layer thickness. The transit velocity is estimated using Ge saturation velocity and i-layer thicknesses of 300 nm and 350 nm for samples A and B, respectively. This limits the photoresponse to $90$ GHz (sample A) and $77$ GHz (sample B). 
For the $f_{RC}$, the series resistance was determined from the forward bias I-V curve to be $120.5\,\Omega$ with a load resistance of $50\,\Omega$. The device capacitance was measured for a $10\,\mu$m-diameter device to be $0.07\,$pF at $5$\,V yielding $f_{RC}=13.34$ GHz.

\medskip
\begin{figure*}[t]
    \centering
    \includegraphics[scale=0.9]{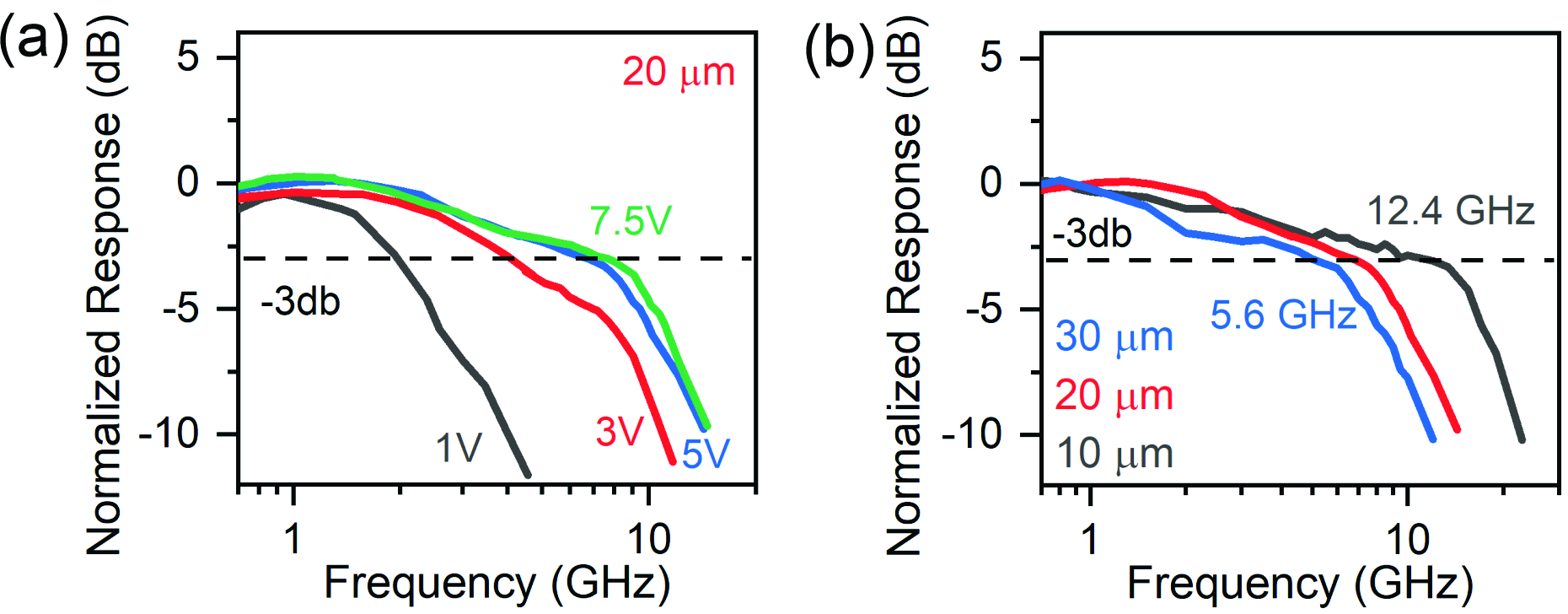}
    \caption{{\bf Photoresponse bandwidth of GeSn PIN PDs.} (a) Normalized photoresponse as a function of the incident optical pulse frequency indicating the photoresponse bandwidth of a $20 \,\mu$m diameter device at various applied biases. (b) same as (a) but showing a comparison of the normalized response for the  $10 \,\mu$m, $20 \,\mu$m, and $30 \,\mu$m diameter devices at $5$ V reverse bias.}
\end{figure*}

\medskip

The normalized responses for both samples are measured, and representative ones are displayed in Fig.3. The normalized response as a function of frequency for the $20 \,\mu$m device at various reverse biases is shown in Fig. 3(a). The measurements were carried out using a $1.55 \,\mu$m single frequency continuous-wave laser that was fiber-coupled to a 40 GHz electro-optical modulator and the output pulsed light was incident on the PD using a fiber focuser lens. To apply DC bias and an RF signal to the PIN PD and to collect the output current, a bias Tee is used along with a parameter analyzer (Keithely 4200A) and a network analyzer (Anritsu 37369D) was used to provide RF signal. The PIN PD output current is connected to a broadband RF amplifier that amplifies the RF signal before it is fed into the network analyzer input. The $-3\,$dB responses values for the $20 \,\mu$m device at biases of 1V, 3V, 5V, and 7.5V were 1.95 GHz, 3.99 GHz, 7.6 GHz, and 7.9 GHz, respectively. Below the RC limit, as the reverse bias increases, the capacitance reduces and the device bandwidth increases. The normalized response of PIN devices with different diameters $30 \,\mu$m, $20 \,\mu$m, and $10 \,\mu$m are compared at 5V, as shown in Fig. 3(b). These devices exhibit a $-3\,$dB responses of 5.6 GHz, 7.6 GHz, and 12.4 GHz, respectively. The $10 \,\mu$m device bandwidth remains below the estimated RC-limited bandwidth value most likely because of the slow diffusion current component contributing to the total device photoresponse.

\noindent {\bf Conclusion}

\noindent This work demonstrates and investigates vertical all-GeSn heterostructures consisting of p-Ge$_{0.92}$Sn$_{0.08}$/i-Ge$_{0.91}$Sn$_{0.09}$/n-Ge$_{0.89}$Sn$_{0.11}$ and p-Ge$_{0.91}$Sn$_{0.09}$/i-Ge$_{0.88}$Sn$_{0.12}$/n-Ge$_{0.87}$Sn$_{0.13}$ heterostructures grown on silicon following a step-graded process. The obtained heterostructures enabled PIN photodetectors exhibiting a high-speed reaching $12.4$ GHz, low noise, and high responsivity across the whole e-SWIR range. The developed all-GeSn heterosturctures offer a viable path to achieve thick, high, and uniform Sn content in the i-layer of the PDs besides a significant relaxation of the compressive strain that is inherent to these epitaxial materials. Consequently, devices based on these all-GeSn heterostructures exhibit a high and stable responsivity over a broader wavelength range. These characteristics highlight the potential of GeSn PIN PDs as effective building blocks for scalable and silicon-compatible e-SWIR technologies.

\noindent {\bf Funding Sources}

\noindent O.M. acknowledges support from NSERC Canada (Discovery, SPG, and CRD Grants), Canada Research Chairs, Canada Foundation for Innovation, Mitacs, PRIMA Qu\'ebec, Defence Canada (Innovation for Defence Excellence and Security, IDEaS), the European Union’s Horizon Europe research and innovation program under grant agreement No 101070700 (MIRAQLS), the US Army Research Office Grant No. W911NF-22-1-0277, and the Air Force Office of Scientific and Research Grant No. FA9550-23-1-0763.

\bigskip

\noindent {\bf Notes}
\noindent The authors declare no conflicts of interest.

\begin{acknowledgement}

The authors thank J. Bouchard for the technical support with the CVD system.

\end{acknowledgement}


\bibliography{main}

\providecommand{\latin}[1]{#1}
\makeatletter
\providecommand{\doi}
  {\begingroup\let\do\@makeother\dospecials
  \catcode`\{=1 \catcode`\}=2 \doi@aux}
\providecommand{\doi@aux}[1]{\endgroup\texttt{#1}}
\makeatother
\providecommand*\mcitethebibliography{\thebibliography}
\csname @ifundefined\endcsname{endmcitethebibliography}  {\let\endmcitethebibliography\endthebibliography}{}
\begin{mcitethebibliography}{45}
\providecommand*\natexlab[1]{#1}
\providecommand*\mciteSetBstSublistMode[1]{}
\providecommand*\mciteSetBstMaxWidthForm[2]{}
\providecommand*\mciteBstWouldAddEndPuncttrue
  {\def\EndOfBibitem{\unskip.}}
\providecommand*\mciteBstWouldAddEndPunctfalse
  {\let\EndOfBibitem\relax}
\providecommand*\mciteSetBstMidEndSepPunct[3]{}
\providecommand*\mciteSetBstSublistLabelBeginEnd[3]{}
\providecommand*\EndOfBibitem{}
\mciteSetBstSublistMode{f}
\mciteSetBstMaxWidthForm{subitem}{(\alph{mcitesubitemcount})}
\mciteSetBstSublistLabelBeginEnd
  {\mcitemaxwidthsubitemform\space}
  {\relax}
  {\relax}

\bibitem[Wun \latin{et~al.}(2016)Wun, Wang, Chen, Bowers, and Shi]{wun2016}
Wun,~J.-M.; Wang,~Y.-W.; Chen,~Y.-H.; Bowers,~J.~E.; Shi,~J.-W. \text{GaSb}-Based PIN Photodiodes With Partially Depleted Absorbers for High-Speed and High-Power Performance at $2.5\,\mu$m Wavelength. \emph{IEEE Trans. Electron Devices} \textbf{2016}, \emph{63}, 2796--2801\relax
\mciteBstWouldAddEndPuncttrue
\mciteSetBstMidEndSepPunct{\mcitedefaultmidpunct}
{\mcitedefaultendpunct}{\mcitedefaultseppunct}\relax
\EndOfBibitem
\bibitem[Williams(2017)]{Williams2017}
Williams,~G.~M. {Optimization of eyesafe avalanche photodiode lidar for automobile safety and autonomous navigation systems}. \emph{Opt. Eng.} \textbf{2017}, \emph{56}, 1 -- 9\relax
\mciteBstWouldAddEndPuncttrue
\mciteSetBstMidEndSepPunct{\mcitedefaultmidpunct}
{\mcitedefaultendpunct}{\mcitedefaultseppunct}\relax
\EndOfBibitem
\bibitem[Kim \latin{et~al.}(2021)Kim, Martins, Jang, Badloe, Khadir, Jung, Kim, Kim, Genevet, and Rho]{Kim2021}
Kim,~I.; Martins,~R.~J.; Jang,~J.; Badloe,~T.; Khadir,~S.; Jung,~H.-Y.; Kim,~H.; Kim,~J.; Genevet,~P.; Rho,~J. Nanophotonics for light detection and ranging technology. \emph{Nat. Nanotechnol.} \textbf{2021}, \emph{16}, 508 -- 524\relax
\mciteBstWouldAddEndPuncttrue
\mciteSetBstMidEndSepPunct{\mcitedefaultmidpunct}
{\mcitedefaultendpunct}{\mcitedefaultseppunct}\relax
\EndOfBibitem
\bibitem[Wang \latin{et~al.}(2021)Wang, Xue, Wan, Zhao, Xu, Liu, Zheng, Zuo, Cheng, and Wang]{wang2021high2}
Wang,~N.; Xue,~C.; Wan,~F.; Zhao,~Y.; Xu,~G.; Liu,~Z.; Zheng,~J.; Zuo,~Y.; Cheng,~B.; Wang,~Q. High-Performance GeSn Photodetector Covering All Telecommunication Bands. \emph{IEEE Photonics J.} \textbf{2021}, \emph{13}, 1--9\relax
\mciteBstWouldAddEndPuncttrue
\mciteSetBstMidEndSepPunct{\mcitedefaultmidpunct}
{\mcitedefaultendpunct}{\mcitedefaultseppunct}\relax
\EndOfBibitem
\bibitem[Li \latin{et~al.}(2021)Li, Peng, Liu, Zhou, Zheng, Xue, Zuo, Chen, and Cheng]{li202130}
Li,~X.; Peng,~L.; Liu,~Z.; Zhou,~Z.; Zheng,~J.; Xue,~C.; Zuo,~Y.; Chen,~B.; Cheng,~B. 30 GHz GeSn photodetector on SOI substrate for 2 $\mu$m wavelength application. \emph{Photonics Res.} \textbf{2021}, \emph{9}, 494--500\relax
\mciteBstWouldAddEndPuncttrue
\mciteSetBstMidEndSepPunct{\mcitedefaultmidpunct}
{\mcitedefaultendpunct}{\mcitedefaultseppunct}\relax
\EndOfBibitem
\bibitem[Razeghi and Nguyen(2014)Razeghi, and Nguyen]{razeghi2014advances}
Razeghi,~M.; Nguyen,~B.-M. Advances in mid-infrared detection and imaging: a key issues review. \emph{Reports on Progress in Physics} \textbf{2014}, \emph{77}, 082401\relax
\mciteBstWouldAddEndPuncttrue
\mciteSetBstMidEndSepPunct{\mcitedefaultmidpunct}
{\mcitedefaultendpunct}{\mcitedefaultseppunct}\relax
\EndOfBibitem
\bibitem[Sieger and Mizaikoff(2016)Sieger, and Mizaikoff]{sieger2016toward}
Sieger,~M.; Mizaikoff,~B. Toward on-chip mid-infrared sensors. \emph{Analytical Chemistry} \textbf{2016}, \emph{88}, 5562--5573\relax
\mciteBstWouldAddEndPuncttrue
\mciteSetBstMidEndSepPunct{\mcitedefaultmidpunct}
{\mcitedefaultendpunct}{\mcitedefaultseppunct}\relax
\EndOfBibitem
\bibitem[Lin \latin{et~al.}(2018)Lin, Luo, Gu, Kimerling, Wada, Agarwal, and Hu]{lin2018mid}
Lin,~H.; Luo,~Z.; Gu,~T.; Kimerling,~L.~C.; Wada,~K.; Agarwal,~A.; Hu,~J. Mid-infrared integrated photonics on silicon: a perspective. \emph{Nanophotonics} \textbf{2018}, \emph{7}, 393--420\relax
\mciteBstWouldAddEndPuncttrue
\mciteSetBstMidEndSepPunct{\mcitedefaultmidpunct}
{\mcitedefaultendpunct}{\mcitedefaultseppunct}\relax
\EndOfBibitem
\bibitem[Popa and Udrea(2019)Popa, and Udrea]{popa2019towards}
Popa,~D.; Udrea,~F. Towards integrated mid-infrared gas sensors. \emph{Sensors} \textbf{2019}, \emph{19}, 2076\relax
\mciteBstWouldAddEndPuncttrue
\mciteSetBstMidEndSepPunct{\mcitedefaultmidpunct}
{\mcitedefaultendpunct}{\mcitedefaultseppunct}\relax
\EndOfBibitem
\bibitem[Chen \latin{et~al.}(2021)Chen, Chen, and Deng]{chen2021recent}
Chen,~B.; Chen,~Y.; Deng,~Z. Recent advances in high speed photodetectors for eSWIR/MWIR/LWIR applications. Photonics. 2021; p~14\relax
\mciteBstWouldAddEndPuncttrue
\mciteSetBstMidEndSepPunct{\mcitedefaultmidpunct}
{\mcitedefaultendpunct}{\mcitedefaultseppunct}\relax
\EndOfBibitem
\bibitem[Ye \latin{et~al.}(2015)Ye, Yang, Gleeson, Pavarelli, Zhang, O'Callaghan, Han, Nudds, Collins, Gocalinska, Pelucchi, O'Brien, Garcia~Gunning, Peters, and Corbett]{Ye2015}
Ye,~N.; Yang,~H.; Gleeson,~M.; Pavarelli,~N.; Zhang,~H.~Y.; O'Callaghan,~J.; Han,~W.; Nudds,~N.; Collins,~S.; Gocalinska,~A.; Pelucchi,~E.; O'Brien,~P.; Garcia~Gunning,~F.~C.; Peters,~F.~H.; Corbett,~B. \text{AlInGaAs} surface normal photodiode for $2\,\mu$m optical communication systems. 2015 IEEE Photonics Conf. (IPC). 2015; pp 456--459\relax
\mciteBstWouldAddEndPuncttrue
\mciteSetBstMidEndSepPunct{\mcitedefaultmidpunct}
{\mcitedefaultendpunct}{\mcitedefaultseppunct}\relax
\EndOfBibitem
\bibitem[Joshi and Becker(2008)Joshi, and Becker]{joshi2008high}
Joshi,~A.; Becker,~D. High-Speed Low-Noise PIN \text{InGaAs} Photoreceiver at $2\,\mu$m Wavelength. \emph{IEEE Photonics Technol. Lett.} \textbf{2008}, \emph{20}, 551--553\relax
\mciteBstWouldAddEndPuncttrue
\mciteSetBstMidEndSepPunct{\mcitedefaultmidpunct}
{\mcitedefaultendpunct}{\mcitedefaultseppunct}\relax
\EndOfBibitem
\bibitem[Yang \latin{et~al.}(2013)Yang, Ye, Phelan, O'Carroll, Kelly, Han, Wang, Nudds, MacSuibhne, Gunning, \latin{et~al.} others]{yang2013}
Yang,~H.; Ye,~N.; Phelan,~R.; O'Carroll,~J.; Kelly,~B.; Han,~W.; Wang,~X.; Nudds,~N.; MacSuibhne,~N.; Gunning,~F.; others Butterfly packaged high-speed and low leakage \text{InGaAs} quantum well photodiode for $2000\,$nm wavelength systems. \emph{Electron. Lett.} \textbf{2013}, \emph{49}, 281--282\relax
\mciteBstWouldAddEndPuncttrue
\mciteSetBstMidEndSepPunct{\mcitedefaultmidpunct}
{\mcitedefaultendpunct}{\mcitedefaultseppunct}\relax
\EndOfBibitem
\bibitem[Chen \latin{et~al.}(2018)Chen, Zhao, Huang, Deng, Cao, Gong, and Chen]{chen2018}
Chen,~Y.; Zhao,~X.; Huang,~J.; Deng,~Z.; Cao,~C.; Gong,~Q.; Chen,~B. Dynamic model and bandwidth characterization of \text{InGaAs/GaAsSb} type-\text{II} quantum wells PIN photodiodes. \emph{Opt. Express} \textbf{2018}, \emph{26}, 35034--35045\relax
\mciteBstWouldAddEndPuncttrue
\mciteSetBstMidEndSepPunct{\mcitedefaultmidpunct}
{\mcitedefaultendpunct}{\mcitedefaultseppunct}\relax
\EndOfBibitem
\bibitem[Tossoun \latin{et~al.}(2018)Tossoun, Zang, Addamane, Balakrishnan, Holmes, and Beling]{tossoun2018}
Tossoun,~B.; Zang,~J.; Addamane,~S.~J.; Balakrishnan,~G.; Holmes,~A.~L.; Beling,~A. \text{InP}-Based Waveguide-Integrated Photodiodes With \text{InGaAs/GaAsSb} Type-\text{II} Quantum Wells and 10-\text{GHz} Bandwidth at $2\,\mu$m Wavelength. \emph{J. Lightwave Technol.} \textbf{2018}, \emph{36}, 4981--4987\relax
\mciteBstWouldAddEndPuncttrue
\mciteSetBstMidEndSepPunct{\mcitedefaultmidpunct}
{\mcitedefaultendpunct}{\mcitedefaultseppunct}\relax
\EndOfBibitem
\bibitem[Chen \latin{et~al.}(2019)Chen, Xie, Huang, Deng, and Chen]{chen2019hi}
Chen,~Y.; Xie,~Z.; Huang,~J.; Deng,~Z.; Chen,~B. High-speed uni-traveling carrier photodiode for 2 $\mu$m wavelength application. \emph{Optica} \textbf{2019}, \emph{6}, 884--889\relax
\mciteBstWouldAddEndPuncttrue
\mciteSetBstMidEndSepPunct{\mcitedefaultmidpunct}
{\mcitedefaultendpunct}{\mcitedefaultseppunct}\relax
\EndOfBibitem
\bibitem[Andreev \latin{et~al.}(2013)Andreev, Serebrennikova, Sokolovskii, Dudelev, Ilynskaya, Konovalov, Kunitsyna, and Yakovlev]{andreev2013high}
Andreev,~I.; Serebrennikova,~O.~Y.; Sokolovskii,~G.; Dudelev,~V.; Ilynskaya,~N.; Konovalov,~G.; Kunitsyna,~E.; Yakovlev,~Y.~P. High-speed photodiodes for the mid-infrared spectral region 1.2\text{--}2.4 $\mu$m based on \text{GaSb/GaInAsSb/GaAlAsSb} heterostructures with a transmission band of 2\text{--}5 \text{GHz}. \emph{Semiconductors} \textbf{2013}, \emph{47}, 1103--1109\relax
\mciteBstWouldAddEndPuncttrue
\mciteSetBstMidEndSepPunct{\mcitedefaultmidpunct}
{\mcitedefaultendpunct}{\mcitedefaultseppunct}\relax
\EndOfBibitem
\bibitem[Moutanabbir \latin{et~al.}(2021)Moutanabbir, Assali, Gong, O'Reilly, Broderick, Marzban, Witzens, Du, Yu, Chelnokov, \latin{et~al.} others]{moutanabbir2021mono}
Moutanabbir,~O.; Assali,~S.; Gong,~X.; O'Reilly,~E.; Broderick,~C.; Marzban,~B.; Witzens,~J.; Du,~W.; Yu,~S.-Q.; Chelnokov,~A.; others Monolithic infrared silicon photonics: the rise of \text{(Si) GeSn} semiconductors. \emph{Appl. Phys. Lett.} \textbf{2021}, \emph{118}, 110502\relax
\mciteBstWouldAddEndPuncttrue
\mciteSetBstMidEndSepPunct{\mcitedefaultmidpunct}
{\mcitedefaultendpunct}{\mcitedefaultseppunct}\relax
\EndOfBibitem
\bibitem[Buca \latin{et~al.}(2022)Buca, Bjelajac, Spirito, Concepci{\'o}n, Gromovyi, Sakat, Lafosse, Ferlazzo, von~den Driesch, Ikonic, \latin{et~al.} others]{buca2022room}
Buca,~D.; Bjelajac,~A.; Spirito,~D.; Concepci{\'o}n,~O.; Gromovyi,~M.; Sakat,~E.; Lafosse,~X.; Ferlazzo,~L.; von~den Driesch,~N.; Ikonic,~Z.; others Room Temperature Lasing in GeSn Microdisks Enabled by Strain Engineering. \emph{Advanced Optical Materials} \textbf{2022}, \emph{10}, 2201024\relax
\mciteBstWouldAddEndPuncttrue
\mciteSetBstMidEndSepPunct{\mcitedefaultmidpunct}
{\mcitedefaultendpunct}{\mcitedefaultseppunct}\relax
\EndOfBibitem
\bibitem[Atalla \latin{et~al.}(2023)Atalla, Kim, Assali, Burt, Nam, and Moutanabbir]{atalla2023extended}
Atalla,~M.~R.; Kim,~Y.; Assali,~S.; Burt,~D.; Nam,~D.; Moutanabbir,~O. Extended-SWIR GeSn LEDs with Reduced Footprint and Efficient Operation Power. \emph{ACS Photonics} \textbf{2023}, \emph{10}, 1649--1653\relax
\mciteBstWouldAddEndPuncttrue
\mciteSetBstMidEndSepPunct{\mcitedefaultmidpunct}
{\mcitedefaultendpunct}{\mcitedefaultseppunct}\relax
\EndOfBibitem
\bibitem[Chang \latin{et~al.}(2022)Chang, Yeh, Jheng, Hsu, Lee, Li, Cheng, and Chang]{chang2022mid}
Chang,~C.-Y.; Yeh,~P.-L.; Jheng,~Y.-T.; Hsu,~L.-Y.; Lee,~K.-C.; Li,~H.; Cheng,~H.; Chang,~G.-E. Mid-infrared resonant light emission from GeSn resonant-cavity surface-emitting LEDs with a lateral PIN structure. \emph{Photonics Research} \textbf{2022}, \emph{10}, 2278--2286\relax
\mciteBstWouldAddEndPuncttrue
\mciteSetBstMidEndSepPunct{\mcitedefaultmidpunct}
{\mcitedefaultendpunct}{\mcitedefaultseppunct}\relax
\EndOfBibitem
\bibitem[Chr{\'e}tien \latin{et~al.}(2022)Chr{\'e}tien, Thai, Frauenrath, Casiez, Chelnokov, Reboud, Hartmann, El~Kurdi, Pauc, and Calvo]{chretien2022room}
Chr{\'e}tien,~J.; Thai,~Q.; Frauenrath,~M.; Casiez,~L.; Chelnokov,~A.; Reboud,~V.; Hartmann,~J.; El~Kurdi,~M.; Pauc,~N.; Calvo,~V. Room temperature optically pumped GeSn microdisk lasers. \emph{Applied Physics Letters} \textbf{2022}, \emph{120}, 051107\relax
\mciteBstWouldAddEndPuncttrue
\mciteSetBstMidEndSepPunct{\mcitedefaultmidpunct}
{\mcitedefaultendpunct}{\mcitedefaultseppunct}\relax
\EndOfBibitem
\bibitem[Elbaz \latin{et~al.}(2020)Elbaz, Buca, von~den Driesch, Pantzas, Patriarche, Zerounian, Herth, Checoury, Sauvage, Sagnes, \latin{et~al.} others]{elbaz2020ultra}
Elbaz,~A.; Buca,~D.; von~den Driesch,~N.; Pantzas,~K.; Patriarche,~G.; Zerounian,~N.; Herth,~E.; Checoury,~X.; Sauvage,~S.; Sagnes,~I.; others Ultra-low-threshold continuous-wave and pulsed lasing in tensile-strained \text{GeSn} alloys. \emph{Nat. Photonics} \textbf{2020}, \emph{14}, 375--382\relax
\mciteBstWouldAddEndPuncttrue
\mciteSetBstMidEndSepPunct{\mcitedefaultmidpunct}
{\mcitedefaultendpunct}{\mcitedefaultseppunct}\relax
\EndOfBibitem
\bibitem[Jung \latin{et~al.}(2022)Jung, Burt, Zhang, Kim, Joo, Chen, Assali, Moutanabbir, Tan, and Nam]{jung2022optically}
Jung,~Y.; Burt,~D.; Zhang,~L.; Kim,~Y.; Joo,~H.-J.; Chen,~M.; Assali,~S.; Moutanabbir,~O.; Tan,~C.~S.; Nam,~D. Optically pumped low-threshold microdisk lasers on a GeSn-on-insulator substrate with reduced defect density. \emph{Photonics Research} \textbf{2022}, \emph{10}, 1332--1337\relax
\mciteBstWouldAddEndPuncttrue
\mciteSetBstMidEndSepPunct{\mcitedefaultmidpunct}
{\mcitedefaultendpunct}{\mcitedefaultseppunct}\relax
\EndOfBibitem
\bibitem[Liu \latin{et~al.}(2022)Liu, Zheng, Niu, Liu, Huang, Li, Zhang, Pang, Liu, Zuo, \latin{et~al.} others]{liu2022sn}
Liu,~X.; Zheng,~J.; Niu,~C.; Liu,~T.; Huang,~Q.; Li,~M.; Zhang,~D.; Pang,~Y.; Liu,~Z.; Zuo,~Y.; others Sn content gradient GeSn with strain controlled for high performance GeSn mid-infrared photodetectors. \emph{Photonics Research} \textbf{2022}, \emph{10}, 1567--1574\relax
\mciteBstWouldAddEndPuncttrue
\mciteSetBstMidEndSepPunct{\mcitedefaultmidpunct}
{\mcitedefaultendpunct}{\mcitedefaultseppunct}\relax
\EndOfBibitem
\bibitem[Luo \latin{et~al.}(2022)Luo, Assali, Atalla, Koelling, Attiaoui, Daligou, Mart{\'\i}, Arbiol, and Moutanabbir]{luo2022extended}
Luo,~L.; Assali,~S.; Atalla,~M.~R.; Koelling,~S.; Attiaoui,~A.; Daligou,~G.; Mart{\'\i},~S.; Arbiol,~J.; Moutanabbir,~O. Extended-SWIR Photodetection in All-Group IV Core/Shell Nanowires. \emph{ACS Photonics} \textbf{2022}, \emph{9}, 914--921\relax
\mciteBstWouldAddEndPuncttrue
\mciteSetBstMidEndSepPunct{\mcitedefaultmidpunct}
{\mcitedefaultendpunct}{\mcitedefaultseppunct}\relax
\EndOfBibitem
\bibitem[Marzban \latin{et~al.}(2022)Marzban, Seidel, Liu, Wu, Kiyek, Zoellner, Ikonic, Schulze, Gr{\"u}tzmacher, Capellini, \latin{et~al.} others]{marzban2022strain}
Marzban,~B.; Seidel,~L.; Liu,~T.; Wu,~K.; Kiyek,~V.; Zoellner,~M.~H.; Ikonic,~Z.; Schulze,~J.; Gr{\"u}tzmacher,~D.; Capellini,~G.; others Strain Engineered Electrically Pumped SiGeSn Microring Lasers on Si. \emph{ACS Photonics} \textbf{2022}, \relax
\mciteBstWouldAddEndPunctfalse
\mciteSetBstMidEndSepPunct{\mcitedefaultmidpunct}
{}{\mcitedefaultseppunct}\relax
\EndOfBibitem
\bibitem[Talamas~Simola \latin{et~al.}(2021)Talamas~Simola, Kiyek, Ballabio, Schlykow, Frigerio, Zucchetti, De~Iacovo, Colace, Yamamoto, Capellini, \latin{et~al.} others]{talamas2021cmos}
Talamas~Simola,~E.; Kiyek,~V.; Ballabio,~A.; Schlykow,~V.; Frigerio,~J.; Zucchetti,~C.; De~Iacovo,~A.; Colace,~L.; Yamamoto,~Y.; Capellini,~G.; others CMOS-compatible bias-tunable dual-band detector based on GeSn/Ge/Si coupled photodiodes. \emph{ACS photonics} \textbf{2021}, \emph{8}, 2166--2173\relax
\mciteBstWouldAddEndPuncttrue
\mciteSetBstMidEndSepPunct{\mcitedefaultmidpunct}
{\mcitedefaultendpunct}{\mcitedefaultseppunct}\relax
\EndOfBibitem
\bibitem[Daligou \latin{et~al.}(2023)Daligou, Soref, Attiaoui, Hossain, Atalla, Del~Vecchio, and Moutanabbir]{daligou2023group}
Daligou,~G.; Soref,~R.; Attiaoui,~A.; Hossain,~J.; Atalla,~M.~R.; Del~Vecchio,~P.; Moutanabbir,~O. Group IV Mid-Infrared Thermophotovoltaic Cells on Silicon. \emph{IEEE Journal of Photovoltaics} \textbf{2023}, \relax
\mciteBstWouldAddEndPunctfalse
\mciteSetBstMidEndSepPunct{\mcitedefaultmidpunct}
{}{\mcitedefaultseppunct}\relax
\EndOfBibitem
\bibitem[Mircovich \latin{et~al.}(2021)Mircovich, Xu, Ringwala, Poweleit, Men{\'e}ndez, and Kouvetakis]{mircovich2021extended}
Mircovich,~M.~A.; Xu,~C.; Ringwala,~D.~A.; Poweleit,~C.~D.; Men{\'e}ndez,~J.; Kouvetakis,~J. Extended Compositional Range for the Synthesis of SWIR and LWIR Ge$_{1-y}$ Sn$_y$ Alloys and Device Structures via CVD of SnH$_4$ and Ge$_3$H$_8$. \emph{ACS Applied Electronic Materials} \textbf{2021}, \emph{3}, 3451--3460\relax
\mciteBstWouldAddEndPuncttrue
\mciteSetBstMidEndSepPunct{\mcitedefaultmidpunct}
{\mcitedefaultendpunct}{\mcitedefaultseppunct}\relax
\EndOfBibitem
\bibitem[Xu \latin{et~al.}(2019)Xu, Ringwala, Wang, Liu, Poweleit, Chang, Zhuang, Men{\'e}ndez, and Kouvetakis]{xu2019synthesis}
Xu,~C.; Ringwala,~D.; Wang,~D.; Liu,~L.; Poweleit,~C.~D.; Chang,~S.~L.; Zhuang,~H.~L.; Men{\'e}ndez,~J.; Kouvetakis,~J. Synthesis and fundamental studies of Si-compatible (Si) GeSn and GeSn mid-IR systems with ultrahigh Sn contents. \emph{Chemistry of Materials} \textbf{2019}, \emph{31}, 9831--9842\relax
\mciteBstWouldAddEndPuncttrue
\mciteSetBstMidEndSepPunct{\mcitedefaultmidpunct}
{\mcitedefaultendpunct}{\mcitedefaultseppunct}\relax
\EndOfBibitem
\bibitem[Soref(2015)]{soref2015enabling}
Soref,~R. Enabling 2 $\mu$m communications. \emph{Nat. Photonics} \textbf{2015}, \emph{9}, 358--359\relax
\mciteBstWouldAddEndPuncttrue
\mciteSetBstMidEndSepPunct{\mcitedefaultmidpunct}
{\mcitedefaultendpunct}{\mcitedefaultseppunct}\relax
\EndOfBibitem
\bibitem[Zhou \latin{et~al.}(2020)Zhou, Miao, Ojo, Tran, Abernathy, Grant, Amoah, Salamo, Du, Liu, \latin{et~al.} others]{zhou2020elec}
Zhou,~Y.; Miao,~Y.; Ojo,~S.; Tran,~H.; Abernathy,~G.; Grant,~J.~M.; Amoah,~S.; Salamo,~G.; Du,~W.; Liu,~J.; others Electrically injected \text{GeSn} lasers on \text{Si} operating up to 100 \text{K}. \emph{Optica} \textbf{2020}, \emph{7}, 924--928\relax
\mciteBstWouldAddEndPuncttrue
\mciteSetBstMidEndSepPunct{\mcitedefaultmidpunct}
{\mcitedefaultendpunct}{\mcitedefaultseppunct}\relax
\EndOfBibitem
\bibitem[Chretien \latin{et~al.}(2019)Chretien, Pauc, Armand~Pilon, Bertrand, Thai, Casiez, Bernier, Dansas, Gergaud, Delamadeleine, \latin{et~al.} others]{chtien2019}
Chretien,~J.; Pauc,~N.; Armand~Pilon,~F.; Bertrand,~M.; Thai,~Q.-M.; Casiez,~L.; Bernier,~N.; Dansas,~H.; Gergaud,~P.; Delamadeleine,~E.; others \text{GeSn} lasers covering a wide wavelength range thanks to uniaxial tensile strain. \emph{ACS Photonics} \textbf{2019}, \emph{6}, 2462--2469\relax
\mciteBstWouldAddEndPuncttrue
\mciteSetBstMidEndSepPunct{\mcitedefaultmidpunct}
{\mcitedefaultendpunct}{\mcitedefaultseppunct}\relax
\EndOfBibitem
\bibitem[Xu \latin{et~al.}(2019)Xu, Wang, Huang, Dong, Masudy-Panah, Wang, Gong, and Yeo]{xu2019high}
Xu,~S.; Wang,~W.; Huang,~Y.-C.; Dong,~Y.; Masudy-Panah,~S.; Wang,~H.; Gong,~X.; Yeo,~Y.-C. High-speed photodetection at two-micron-wavelength: technology enablement by \text{GeSn/Ge} multiple-quantum-well photodiode on 300 mm \text{Si} substrate. \emph{Opt. Express} \textbf{2019}, \emph{27}, 5798--5813\relax
\mciteBstWouldAddEndPuncttrue
\mciteSetBstMidEndSepPunct{\mcitedefaultmidpunct}
{\mcitedefaultendpunct}{\mcitedefaultseppunct}\relax
\EndOfBibitem
\bibitem[Tran \latin{et~al.}(2019)Tran, Littlejohns, Thomson, Pham, Ghetmiri, Mosleh, Margetis, Tolle, Mashanovich, Du, \latin{et~al.} others]{tran2019study}
Tran,~H.; Littlejohns,~C.~G.; Thomson,~D.~J.; Pham,~T.; Ghetmiri,~A.; Mosleh,~A.; Margetis,~J.; Tolle,~J.; Mashanovich,~G.~Z.; Du,~W.; others Study of \text{GeSn} mid-infrared photodetectors for high frequency applications. \emph{Front. in Mater.} \textbf{2019}, \emph{6}, 278\relax
\mciteBstWouldAddEndPuncttrue
\mciteSetBstMidEndSepPunct{\mcitedefaultmidpunct}
{\mcitedefaultendpunct}{\mcitedefaultseppunct}\relax
\EndOfBibitem
\bibitem[Talamas~Simola \latin{et~al.}(2021)Talamas~Simola, Kiyek, Ballabio, Schlykow, Frigerio, Zucchetti, De~Iacovo, Colace, Yamamoto, Capellini, \latin{et~al.} others]{talamas2021}
Talamas~Simola,~E.; Kiyek,~V.; Ballabio,~A.; Schlykow,~V.; Frigerio,~J.; Zucchetti,~C.; De~Iacovo,~A.; Colace,~L.; Yamamoto,~Y.; Capellini,~G.; others \text{CMOS}-Compatible Bias-Tunable Dual-Band Detector Based on \text{GeSn/Ge/Si} Coupled Photodiodes. \emph{ACS Photonics} \textbf{2021}, \emph{8}, 2166--2173\relax
\mciteBstWouldAddEndPuncttrue
\mciteSetBstMidEndSepPunct{\mcitedefaultmidpunct}
{\mcitedefaultendpunct}{\mcitedefaultseppunct}\relax
\EndOfBibitem
\bibitem[Tsai \latin{et~al.}(2021)Tsai, Lin, Cheng, Lee, Cheng, and Chang]{tsai2021gesn}
Tsai,~C.-H.; Lin,~K.-C.; Cheng,~C.-Y.; Lee,~K.-C.; Cheng,~H.; Chang,~G.-E. GeSn lateral pin waveguide photodetectors for mid-infrared integrated photonics. \emph{Optics letters} \textbf{2021}, \emph{46}, 864--867\relax
\mciteBstWouldAddEndPuncttrue
\mciteSetBstMidEndSepPunct{\mcitedefaultmidpunct}
{\mcitedefaultendpunct}{\mcitedefaultseppunct}\relax
\EndOfBibitem
\bibitem[Cui \latin{et~al.}(2023)Cui, Zheng, Zhu, Liu, Huang, Liu, Zuo, and Cheng]{cui2023sn}
Cui,~J.; Zheng,~J.; Zhu,~Y.; Liu,~X.; Huang,~Q.; Liu,~Z.; Zuo,~Y.; Cheng,~B. Sn component gradient GeSn photodetector with 3 dB bandwidth over 50 GHz for extending L band telecommunication. \emph{Optics Letters} \textbf{2023}, \emph{48}, 6148--6151\relax
\mciteBstWouldAddEndPuncttrue
\mciteSetBstMidEndSepPunct{\mcitedefaultmidpunct}
{\mcitedefaultendpunct}{\mcitedefaultseppunct}\relax
\EndOfBibitem
\bibitem[Tran \latin{et~al.}(2019)Tran, Pham, Margetis, Zhou, Dou, Grant, Grant, Al-Kabi, Sun, Soref, \latin{et~al.} others]{tran2019si}
Tran,~H.; Pham,~T.; Margetis,~J.; Zhou,~Y.; Dou,~W.; Grant,~P.~C.; Grant,~J.~M.; Al-Kabi,~S.; Sun,~G.; Soref,~R.~A.; others Si-based \text{GeSn} photodetectors toward mid-infrared imaging applications. \emph{ACS Photonics} \textbf{2019}, \emph{6}, 2807--2815\relax
\mciteBstWouldAddEndPuncttrue
\mciteSetBstMidEndSepPunct{\mcitedefaultmidpunct}
{\mcitedefaultendpunct}{\mcitedefaultseppunct}\relax
\EndOfBibitem
\bibitem[Atalla \latin{et~al.}(2022)Atalla, Assali, Koelling, Attiaoui, and Moutanabbir]{atalla2022high}
Atalla,~M.~R.; Assali,~S.; Koelling,~S.; Attiaoui,~A.; Moutanabbir,~O. High-bandwidth extended-SWIR GeSn photodetectors on silicon achieving ultrafast broadband spectroscopic response. \emph{ACS Photonics} \textbf{2022}, \emph{9}, 1425--1433\relax
\mciteBstWouldAddEndPuncttrue
\mciteSetBstMidEndSepPunct{\mcitedefaultmidpunct}
{\mcitedefaultendpunct}{\mcitedefaultseppunct}\relax
\EndOfBibitem
\bibitem[Atalla \latin{et~al.}(2021)Atalla, Assali, Attiaoui, Lemieux-Leduc, Kumar, Abdi, and Moutanabbir]{atalla2021all}
Atalla,~M.~R.; Assali,~S.; Attiaoui,~A.; Lemieux-Leduc,~C.; Kumar,~A.; Abdi,~S.; Moutanabbir,~O. All-Group \text{IV} Transferable Membrane Mid-Infrared Photodetectors. \emph{Adv. Funct. Mater.} \textbf{2021}, \emph{31}, 2006329\relax
\mciteBstWouldAddEndPuncttrue
\mciteSetBstMidEndSepPunct{\mcitedefaultmidpunct}
{\mcitedefaultendpunct}{\mcitedefaultseppunct}\relax
\EndOfBibitem
\bibitem[Dong \latin{et~al.}(2017)Dong, Wang, Xu, Lei, Gong, Guo, Wang, Lee, Loke, Yoon, \latin{et~al.} others]{dong2017two}
Dong,~Y.; Wang,~W.; Xu,~S.; Lei,~D.; Gong,~X.; Guo,~X.; Wang,~H.; Lee,~S.-Y.; Loke,~W.-K.; Yoon,~S.-F.; others Two-micron-wavelength germanium-tin photodiodes with low dark current and gigahertz bandwidth. \emph{Opt. Express} \textbf{2017}, \emph{25}, 15818--15827\relax
\mciteBstWouldAddEndPuncttrue
\mciteSetBstMidEndSepPunct{\mcitedefaultmidpunct}
{\mcitedefaultendpunct}{\mcitedefaultseppunct}\relax
\EndOfBibitem
\bibitem[Lin \latin{et~al.}(2017)Lin, Lee, Bao, Guo, Wang, Michel, and Tan]{lin2017high}
Lin,~Y.; Lee,~K.~H.; Bao,~S.; Guo,~X.; Wang,~H.; Michel,~J.; Tan,~C.~S. High-efficiency normal-incidence vertical pin photodetectors on a germanium-on-insulator platform. \emph{Photonics Res.} \textbf{2017}, \emph{5}, 702--709\relax
\mciteBstWouldAddEndPuncttrue
\mciteSetBstMidEndSepPunct{\mcitedefaultmidpunct}
{\mcitedefaultendpunct}{\mcitedefaultseppunct}\relax
\EndOfBibitem
\end{mcitethebibliography}

%\begin{figure}[htbp]
%\centering
%\fbox{\includegraphics[width=\textwidth]{TOC_final2.jpg}}
%\caption{Table of contents figure.}
%\label{fig:Fig__SI}
%\end{figure}

\end{document}